\documentclass[conference]{IEEEtran}
\IEEEoverridecommandlockouts
\usepackage{caption}

\usepackage{cite}
\usepackage{amsmath,amssymb,amsfonts}
\usepackage{algorithmic}
\usepackage{graphicx}
\usepackage{textcomp}
\usepackage{xcolor}
\usepackage{svg}
\usepackage{amsmath}
\usepackage{url}
\usepackage{lettrine}
\usepackage{graphicx}

\def\BibTeX{{\rm B\kern-.05em{\sc i\kern-.025em b}\kern-.08em
    T\kern-.1667em\lower.7ex\hbox{E}\kern-.125emX}}

\begin{document}
\title{BlockLot: Blockchain-based Verifiable Lottery 
}

\author{\IEEEauthorblockN{
Yongrae Jo\IEEEauthorrefmark{1},
and Chanik Park  \IEEEauthorrefmark{2}
}
\IEEEauthorblockA{Department of Computer Science and Engineering \\
Pohang University of Science and Technology \\
\IEEEauthorrefmark{1}memex@postech.ac.kr,
\IEEEauthorrefmark{2}cipark@postech.ac.kr
}}

\maketitle
\thispagestyle{plain}
\pagestyle{plain}

\begin{abstract}
We propose BlockLot, a blockchain based verifiable lottery. BlockLot provides transparent, immutable, fair, and verifiable lottery services enhanced by recent blockchain technologies such as append-only (replicated) distributed ledger and smart contract. In addition, BlockLot allows all participants to perform various verification procedures to ensure that the system is actually working as expected. 


We implement BlockLot services on Hyperledger/Fabric blockchain platform with web-based user interface. 
\end{abstract}


\section{Introduction}
\IEEEPARstart{L}{ottery} events occur frequently. A lottery event requires trust between participants and event organizers. In general, people participate in the event because they believe the probability of winning lottery is fair. But the trust between the two is sometimes broken. For example, an event organizer can manipulate the results of the winner.

To solve the problem of fraudulent lottery, computer software can be used to compute winners without human intervention. Lottery program, which is determined solely by input and output, is believed to be more reliable, fair, transparent and verifiable than human-involved one.

The existing lottery systems adopt a centralized approach, in which calculations are performed on a single computer\cite{unipicker,kindergarden,zhou2001playing,goldschlag1998publicly}. However, in such a case, a person who gains privileged access to the computer (e.g., Hacker), on which the lottery program is run, may harm its integrity, transparency, and fairness. For example, an attacker who gains privileged access can manipulate the winners after the results are issued. Therefore, the centralized lottery system is unreliable in nature.

We define three attributes of the unreliable lottery which includes predictability, modifiability and information hiding. Predictability means that the random seed used in the lottery can be predicted. Modifiability means manipulation of event information, lottery result, random seed, lottery algorithm. Information hiding represents hiding important details of the lottery (e.g., changing a prize). 

Against the attributes of unreliable lottery system, we also defined four attributes of reliable lottery which includes fairness, transparency, immutability, and verifiability. Fairness means that the winner should not be predictable in advance. Transparency indicates that the method of drawing winners and the results of the draw should be transparent to all participants. Immutability means that the registered information can not be modified at one’s disposal and no one can change the result after it is announced. Verifiability represents that the lottery system should satisfy the aforementioned attributes and that the results should be verifiable.

We introduce BlockLot, a reliable lottery system based on blockchain. BlockLot enables reliable lottery in a way of fair, transparent, immutable, and verifiable. BlockLot uses a Bitcoin blockchain as a random beacon to choose winners, which it is unpredictable and publicly verifiable\cite{bentov2016bitcoin, bonneau2015bitcoin}. BlockLot also uses a distributed-replicated ledger based on a blockchain and all services in BlockLot are supported by smart contracts in distributed manner. The lottery services provided by BlockLot includes \emph{open}, \emph{query}, \emph{subscribe}, \emph{draw}, \emph{check}, and \emph{verify}. The event organizer can open a lottery by providing required information such as the announcement date, number of winners, and list of prizes. Participants can query the lottery events registered and subscribe to the event to participate in it. When conditions for drawing winners are satisfied, an event organizer draw winners and participants can check and verify it.


The paper is organized as follows. Section \ref{sec:designandimplementation} describes design and implementation of BlockLot. Section \ref{sec:discussionandrelatedwork} presents discussion and related work. Section \ref{sec:conclusion} includes concluding remarks.

\section{BlockLot}
\label{sec:designandimplementation}

\subsection{Overview}
\begin{figure}
    \centering
    \includegraphics[]{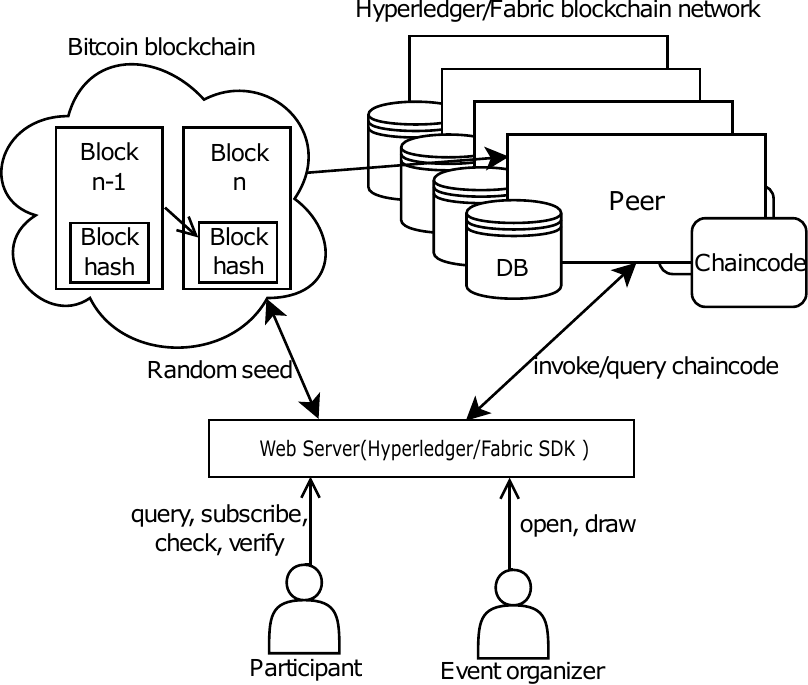}
    \caption{Overview of BlockLot: Users can use BlockLot services through a web server. The peer nodes in the back-end blockchain network maintain the lottery ledger and execute transactions of lottery service. Peer uses Bitcoin block hash for darwing winners.}
    \label{fig:arch}
\end{figure}

We developed BlockLot on Hyperledger/Fabric, a popular permissioned blockchain platform for enterprise\cite{fabric}. Hyperledger/Fabric uses peer nodes to maintain a distributed ledger and execute chaincodes(a.k.a smart contracts). The Web server and the Hyperledger/Fabric SDK interact with the user (web client) and the blockchain network. Upon receiving the user's transaction, the web server sends it to the peer node and informs the user of the chaincode execution result. Also, it fetches the target block number to be used as a random seed from the Bitcoin random beacon. (Fig. \ref{fig:arch})

\begin{figure}
    \centering
    \includegraphics[scale=0.2]{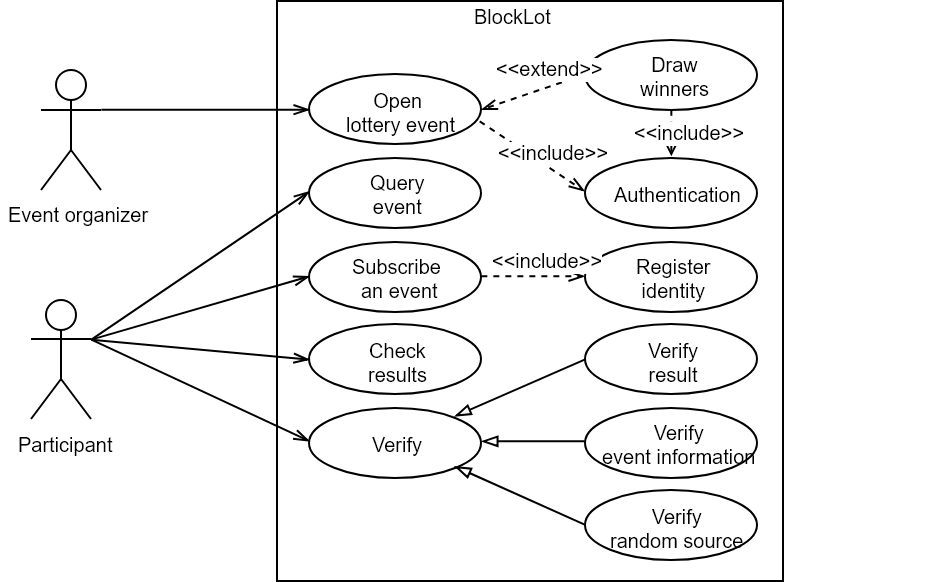}
    \caption{Usecase diagram of BlockLot: Event organizer can open a lottery event and draw winners. Participant can query events, subscribe to an event, check and verify the result.}
    \label{fig:usercase}
\end{figure}

BlockLot users can be divided into participants and event organizers. An event organizer can register the lottery event and draw the winner with authentication (Fig. \ref{fig:usercase}). Participants can query the event, subscribe to the lottery, and check and verify the lottery result.

Chaincode is a special program that processes transactions and performs I/O with blockchain database in Hyperledger/Fabric. The types of transaction of BlockLot consists of query, open, draw, and subscribe, each of which is implemented by chaincode. We used GoLang to write chaincode because Hyperledger/Fabric supports GoLang as a primary language. 

\subsection{Generating a randomness}

\subsection{Lottery sequence}
\begin{figure}
    \centering
    \includegraphics[scale=0.2]{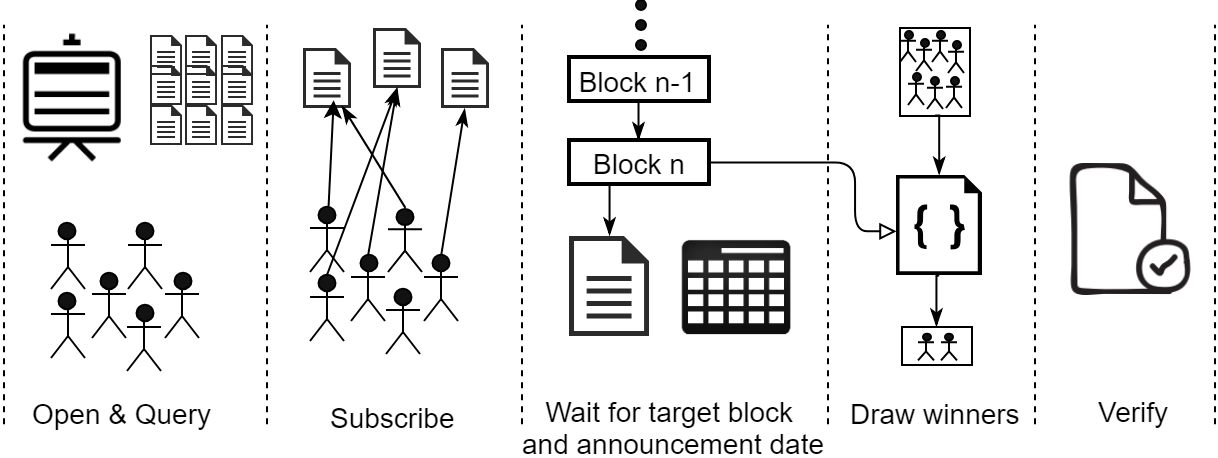}
    \caption{Lottery sequence: 1) When an event organizer opens a lottery, participants can query the event, and 2) subscribe to the event. 3) After the conditions for drawing winners are satisfied, 4) the event organizer draw winners, and the participants can check and verify the results.}
    \label{fig:sequence}
\end{figure}

In this section, we introduce the detailed implementation of the lottery service of BlockLot (Fig. \ref{fig:sequence}). 

\subsubsection{Open}
A open transaction registers a lottery in the blockchain. An event organizer provides event information such as name, announcement date, number of winners, block offset. The block offset represents the offset from the most recent block height in the main chain of the Bitcoin blockchain at the time of registration(namely target block). We used developer API provided by \cite{blockchaininfo}. During the process, a unique ID of the event is generated and inserted into the key/value store of ledger with related information.

\subsubsection{Query}
A query transaction queries all registered events in the blockchain. We use the range query function \textsf{GetStateByRange()}, which is provided by the chain code shim interface, to retrieve the strings of (key, value) pairs and display the concatenated values to the participant.

\subsubsection{Subscribe}
To subscribe to an event, a participant chooses an event, enters their identity(e.g., email address) and gets a authentication token. The token is used to prove that the participant is actually a winner after the drawing. Subscribe transaction appends a participant to the list of participants of given event. The hash value of the registration name concatenated by authentication token is recorded on the blockchain, ensuring the anonymity of the participant.

\subsubsection{Draw}
A draw transaction can be invoked after the announcement date has arrived and a target block has been published. The event organizer enters a pre-issued authentication token, and if the verification is passed, winners will be selected. Once the winners are selected, all the information about the lottery event is fixed, so the information obtained by using the hash function is used as a \textit{verifiable key}.(Fig. \ref{fig:verifiablekey}). The winner list and the verifiable key are stored in the blockchain. The script for draw uses the \textit{Fisher-Yates random shuffle algorithm}, which uses the block hash of the target block as a random seed, to output the winner array in a deterministic way.

\subsubsection{Check}
After the draw transaction has been completed, the participant can confirm the winner by entering the name and authentication token. Like the participant list, the winner list is also expressed in hash values.

\subsubsection{Verify}
Participants also need to verify the result to ensure that the lottery is conducted without compromising the attributes of a  reliable lottery. Our assumptions on an attacker contains a) predicting and manipulating random seed, b) fabricating event information arbitrarily, c) Manipulating winner list after draw.(i.e., recording a fraudulent winner list), d) malicious lottery scripts(e.g., draw winners regardless of random seed).

\paragraph{Preventing an attack on random seed and verifying its integrity} We use Bitcoin block hash, which is inherently unpredictable and immutable. Bitcoin miners spread around the world compete to get reward by solving random-based hash puzzles, so they generate unpredictable random numbers. Also, Bitcoin is the largest global blockchain at the moment, which it is immune to control  block hash. Using Bitcoin blockchain as randomness source is described in \cite{bonneau2015bitcoin, bentov2016bitcoin}. In Bitcoin, block hash is calculated using its \emph{version, previousHash, merkle\_root, timestamp, bits, nonce}  fields of the block. So we calculate the hash value of the block and compare it with the one in the lottery in BlockLot.

\paragraph{Verifying event information integrity} We also recalculate and compare hash values with information determined at the time of event registration and draw to verify integrity.  
Given the participant list, random seed, and draw script, participant can calculate the winner list and compare it to the winner list on the blockchain. 

\paragraph{Verifying response results from multiple peers} A peer that holds a blockchain can be crashed or hacked, so we need to query all peers in the network, and compare the response results. If more than half of the peers return the same response results, users can accept it. 

\paragraph{Verifying statistical fairness} The draw script is verified by z-test to verify that a particular participant does not win more than a average number of times. Given $n$ times of running lottery, and when the probability of winning lottery is $p$ (number of winners / number of participants), its binomial distribution should follow a normal distribution if $n$ is large enough. If each participant's z-score is less than a pre-defined maximum z-value, the verification is successful, otherwise it fails.

\begin{figure}
    \centering
    \includegraphics[scale=0.55]{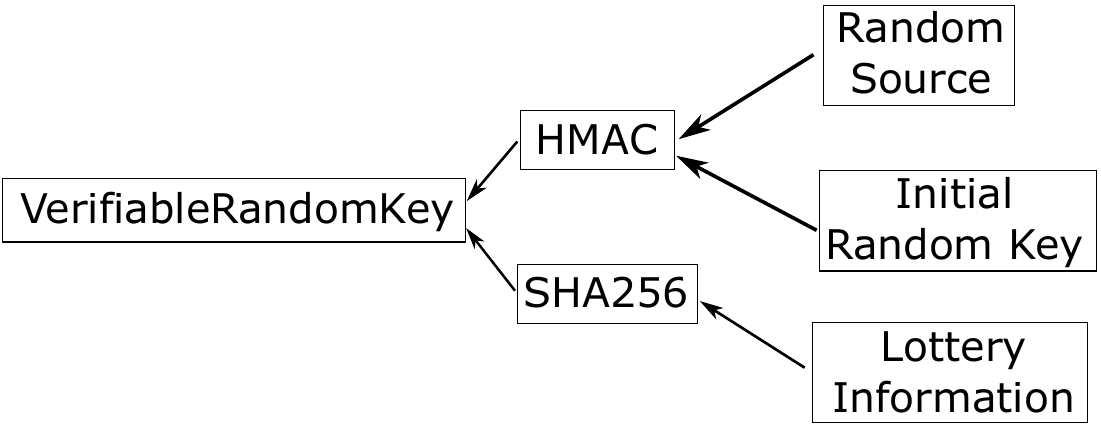}
    \caption{Deriving verifiable random key: The key is derived from concatenating the result from HAMC(initital random key, random source) and SHA256(Lottery informatino)}
    \label{fig:verifiablekey}
\end{figure}


\section{Implementation}

\subsection{Chaincode Implementation}
Hereby we explain how to handle lottery transactions with chaincode implementation in Hyperledger/Fabric. Shim package provides simple interface to access and modify ledger. For example, \textsf{GetState()} is used to access to the ledger and \textsf{PutState()} is used to modify it. Unless otherwise stated, all of the functions below are included in the stub interface.


\begin{figure}
    \centering
    \includegraphics[scale=0.2]{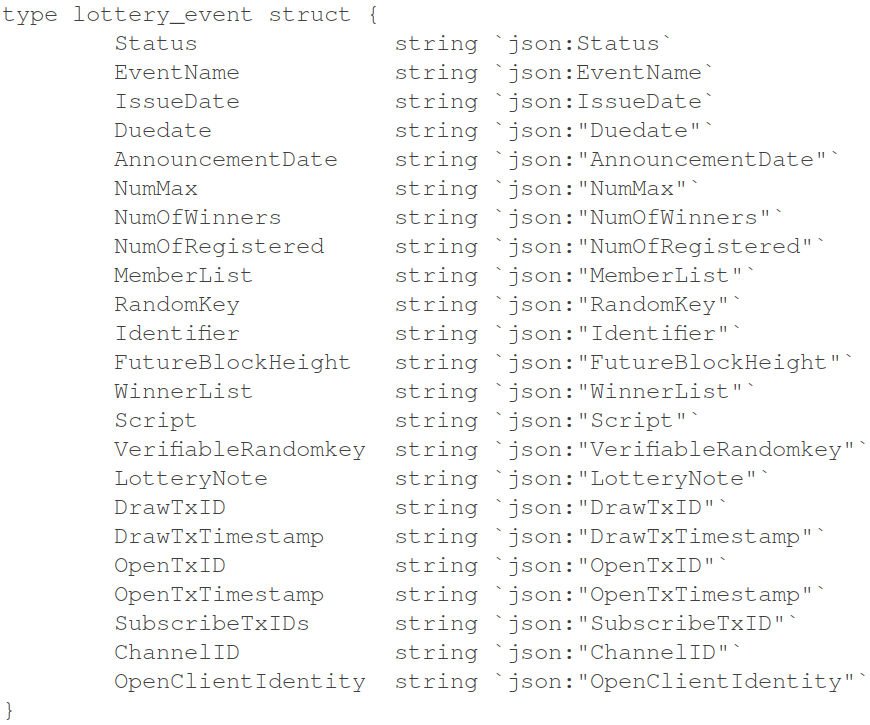}
    \caption{Lottery event representation in chaincode}
    \label{fig:struct}
\end{figure}

\subsubsection{Open}
Implementing an \textit{open} transaction is straightforward. It creates a new lottery\_event struct, fills fields in the struct, marshals the struct, and inserts it into a ledger using \textsf{PutState()} with its identifier as a key. Since the ledger is maintained per channel in Hyperledger/Fabric, \textit{ChannelID} should be specified when the event is created. The ID of \textit{open} transaction is recorded as well. \textsf{GetChannel()} and \textsf{GetTxID()} is used to do these. \textit{DrawTxID} and \textit{SubscribeTxIDs} are initialized as UNDEFINED. Status field is initialized as REGISTERED.

\subsubsection{Query}
We used \textsf{GetStateByRange()} to iterate all registered lottery events in the ledger. By iterating all keys in the ledger, we can combine those key/value pairs into one byte array. The byte array is returned to client.

\subsubsection{Subscribe}
A \textit{subscribe} transaction contains identifiers of event and member as function arguments. It retrieves an event using \textsf{GetState()}, checks if the member is already registered, compares timeliness(i.e., due date). If successful, it adds the member to \textit{MemberList} in lottery event. After that, it also records \textit{subscribeTxID} returned from \textsf{GetTxID()}.

\subsubsection{Draw}
It fetches target block from \cite{blockchaininfo}, extracts hash value from it, and uses it as a random seed for the Fisher-Yates random shuffle algorithm\cite{fisher1943statistical}. We implement the algorithm in GoLang. The algorithm receives the number of participants$\mathsf{P}$, number of winners($\mathsf{W}$), and random source as arguments. Since the algorithm requires a fresh random seed in each iteration step to swap two elements in an array with size of $\mathsf{P}$, we newly design \textsf{randomOracle()} to generate new random values from the random source(i.e., block hash). \textsf{randomOracle()} simply hashes the random source and extracts the first four 32-bit integers from the hashed value, and adds them together to get a single value. Then, the result value is used to choose two elements randomly in the array. After iterating the array by $\mathsf{P}$ times, the algorithm is terminated. As a result, first $\mathsf{W}$ elements from the array are selected as winners.

Draw also uses two cryptographic hash functions to create \textit{VerifiableRandomKey}. The key is derived from HMAC with random source, initial random key, and SHA256 with lottery information. (Fig. \ref{fig:verifiablekey}).
The initial random key is issued when the event organizer register a lottery event. Lottery information includes all fields in lottery event struct (Fig. \ref{fig:struct}). Since the winner list is determined, no information in the lottery event is changed. Therefore, the key for verifying integrity can be derived from these values.


\subsection{User Interface}
\begin{figure}
    \centering
    \includegraphics[scale=0.2]{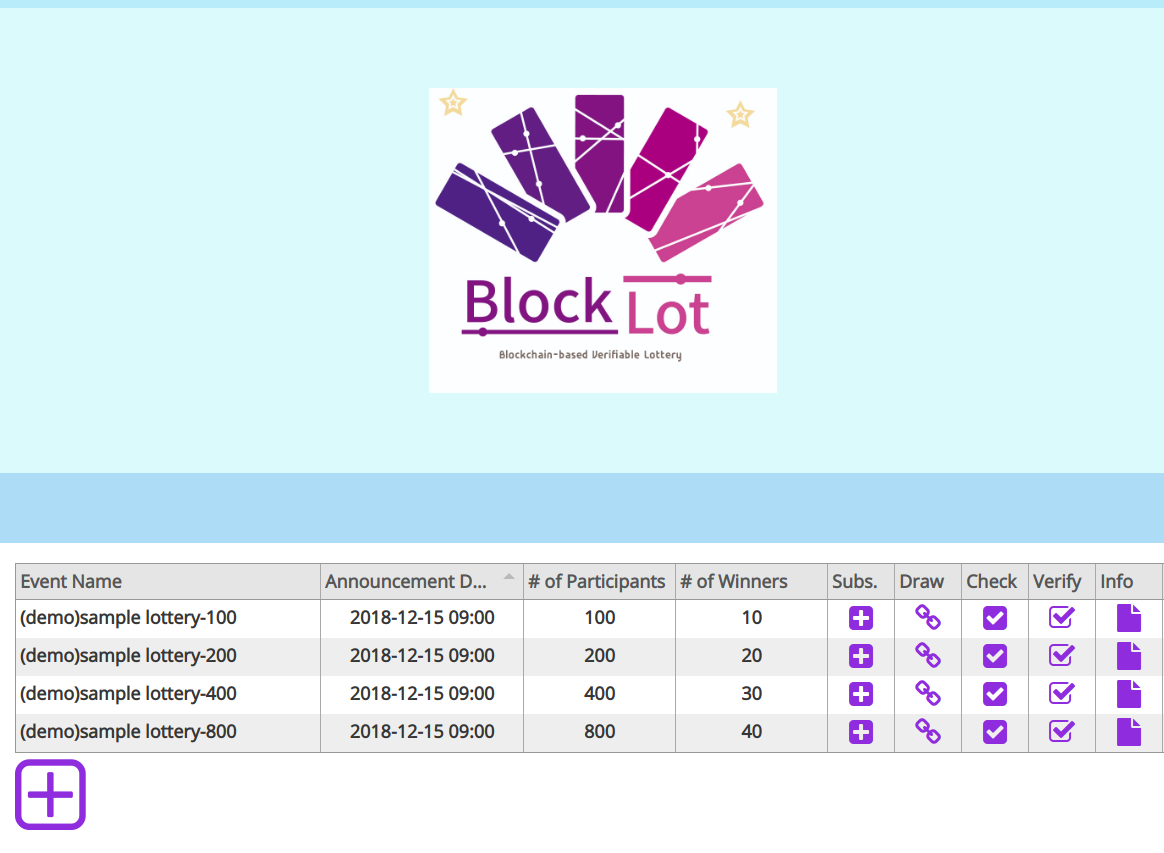}
    \caption{Main user interface: BlockLot services are available via web browser based interface. Lotteries are displayd as tables. Users can use each services by clicking the corresponding buttons. For example, a user can open a lottery by clicking the big button with a plus sign at the bottom of the table.}
    \label{fig:user_interface}
\end{figure}

Main user interface for BlockLot is implemented in web browser (Fig. \ref{fig:user_interface}). Lotteries are displayed as tables, which contain event name, announcement date, number of participants, number of winners, buttons for subscribe, draw, check, verify, and info.

\begin{figure}
    \centering
    \includegraphics[scale=0.3]{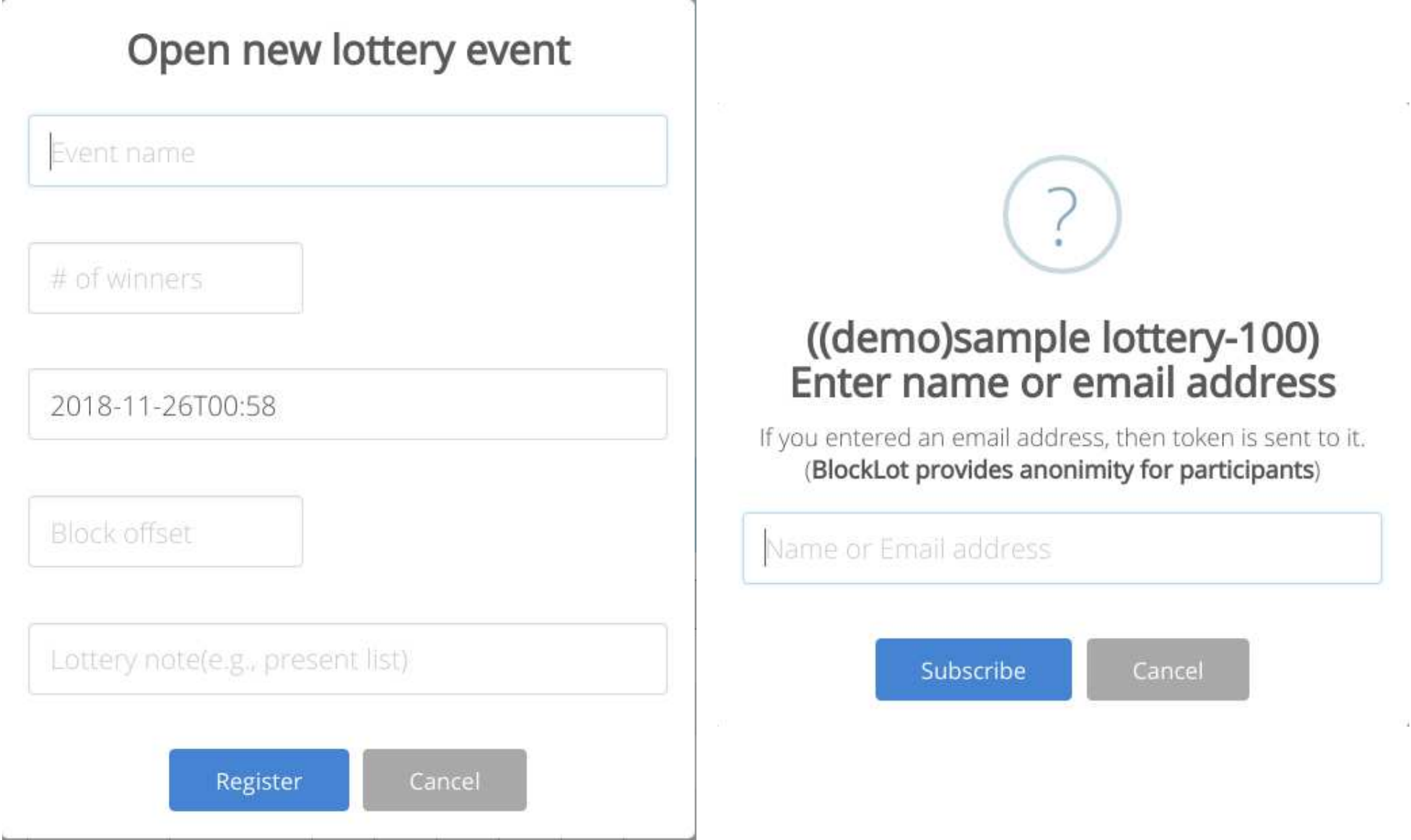}
    \caption{Open \& Subscribe Interface: An event organizer can open a lottery event by providing its name, number of winners, announcement date, block offset, and note. Participants participate in a lottery event by providing his/her name or e-mail address.}
    \label{fig:Open_Subscribe}
\end{figure}

An event organizer can register the lottery event by clicking the big button with a plus sign at the bottom of the table. A user can participate in a lottery event by clicking the subs. button and providing his/her name or email address (Fig. \ref{fig:Open_Subscribe}). After that, an authentication token will be issued to prove that the user is a winner when it actually is.

We implemented full-fledged BlockLot and it is publicly available on \cite{BlockLot}. More implementation details as well as user interfaces that is not covered here can be found on the github repository.

\section{Discussion and Related work}
\label{sec:discussionandrelatedwork}

\subsection{Security}
\paragraph{Anonymity} BlockLot assumes that the participants identities are known offline in prior, meaning that their privacy may not be strictly guaranteed. For example, participation information to an event or the final results can be publicly available. We believe that this assumption is not a big problem because many lottery events provided by an enterprise in real-world often require participant's identity. (e.g., customer’s name).

\paragraph{Duplicate participation} Current implementation does not strictly prevent one user from participating in the same lottery multiple times. Because the user authentication is done only by name or e-mail, a malicious user can easily generate multiple identities to increase the chances of winning lottery. However, this concern will not be a problem if participant's identities are known offline to the event organizer.

\paragraph{Lagged blockchain view} We used developer APIs from \cite{blockchaininfo}, a freely available blockchain service provider to read the latest Bitcoin block. One significant problem here is that the main server which receives blocks on that site sometimes see a lagged blockchain view (e.g., temporary network delay). In that case, an attacker can predict the random seed, and exploit this information to be winner. But this limitation can be relaxed by choosing a target block which is much higher than currently perceived latest block.

\subsection{Deployment environment}
BlockLot focuses on lottery in situations where participants' identities are known in advance. For example, when a company performs a lottery for promotional purposes, the customer provides his/her identity to participate in the lottery. 
\subsection{Related work}
In traditional centralized lottery system, a lottery is executed in a single computer. But this approach is vulnerable to a single point of failure as well as to the existence of a hacker who can gain privileged access to the system. Our work is basically a decentralized system, which means that a single participant cannot manipulate any data on the blockchain.

Also, blockchain-based lottery systems such as Kibo\cite{kibo}, FireLotto\cite{firelotto}, Quanta\cite{quanta}, and SmartBilions\cite{smartbillions} are based on Ethereum blockchain\cite{ethereum}. Ethereum provides a turing-complete smart contract platform which is suitable for executing lottery services. In contrast,  our system is based on a permission-based blockchain, mainly for enterprise, which is orthogonal to Ethereum-based methods because our system requires identity to be known to participate in a lottery.

\section{Conclusion}
\label{sec:conclusion}
In this paper, we present BlockLot, a Blockchain based verifiable lottery that records all information about the lottery event in a distributed ledger, and conducts the event in a transparent and immutable manner. BlockLot uses Bitcoin block hash as a unpredictable(i.e., fairness) random seed to draw winners. BlockLot also offers a variety of verification methods to ensure that the lottery is actually reliable. In conclusion, BlockLot is expected to significantly reduce social costs due to fraudulent lotteries.

\section*{Acknowledgment}
This work was partially supported by The Institute for Information \& Communications Technology Promotion (IITP) in South Korea, 2018, funded by the Korea government (Ministry of Science and ICT) (No.2017-0-00652, Development of High-performance and High-reliable Blockchain for Distributed Autonomous IoT Platform). 
We also would like to thank to undergraduate students in POSTECH. Hyunseung Lee gave supplementary comments on this original document, Byoungyun Park helped with testing BlockLot, and Minsoo Koo designed BlockLot logo.


\bibliography{main}{}
\bibliographystyle{plain}

\end{document}